\documentclass[showpacs,preprintnumbers,amsmath,amssymb]{revtex4}
\usepackage{graphicx}
\usepackage{dcolumn}
\usepackage{bm}

\begin{document}
\def\bsv{{\hbox{\boldmath $v$}}}
\def\bsk{{\hbox{\boldmath $k$}}}

\title{Signature of Granular Structures by Single-Event Intensity
Interferometry }

\author{Cheuk-Yin Wong$^{1,2}$}
\author{Wei-Ning Zhang$^{3}$}

\affiliation{
$^1$Physics Division, Oak Ridge National Laboratory, Oak Ridge, TN
37831, U.S.A.\\
$^2$Department of Physics, University of Tennessee, Knoxville, TN
37996, U.S.A.\\
$^3$Department of Physics, Harbin Institute of Technology, 
Harbin, 150006, P. R. China
}

\date{\today}

\begin{abstract}
The observation of a granular structure in high-energy heavy-ion
collisions can be used as a signature for the quark-gluon plasma phase
transition, if the phase transition is first order in nature.  We
propose methods to detect a granular structure by the single-event
intensity interferometry. We find that the correlation function from a
chaotic source of granular droplets exhibits large fluctuations, with
maxima and minima at relative momenta which depend on the relative
coordinates of the droplet centers.  The presence of this type of
maxima and minima of a single-event correlation function at many
relative momenta is a signature for a granular structure and a
first-order QCD phase transition.  We further observe that the Fourier
transform of the correlation function of a granular structure exhibits
maxima at the relative spatial coordinates of the droplet centers,
which can provide another signature of the granular structure.
\end{abstract}

\pacs{25.75.-q, 25.75.Nq, 25.75.Gz}

\maketitle

\section{Introduction and Summary}

Recently, much progress has been made in the experimental search for
the quark-gluon plasma \cite{Gyu04,Rik04}.  The occurrence of
jet-quenching and collective flow suggests the presence of a very
dense matter produced in high-energy heavy-ion collisions
\cite{Gyu04,Rik04}.  A very important question is whether the produced
dense matter is a quark-gluon plasma.  If it is a quark-gluon plasma,
it will undergo a phase transition from the quark-gluon plasma phase
to the hadronic phase.  It is of great interest to search for the
signature for the phase transition of the quark-gluon plasma.

The signature for the phase transition depends sensitively on the
order of the transition.  Previously, Witten and many other workers
noted that a granular structure of droplets occurs in a first-order
QCD phase transition, and the observation of the granular structure can
be used as a signature for a first-order QCD phase transition
\cite{Wit84}-\cite{Ran04a}.  In a recent
spinodal analysis, Randrup found that the spinodal instability has a
high degree of amplification and the instability of the most rapidly
amplified wavelengths grow predominantly, leading to spinodal
(granular) density patterns that may indeed be used as a diagnostic
tool for a first-order QCD phase transition \cite{Ran04}.  Assuming
that the granular particles are arranged in regular rapidity intervals
in momentum space, Randrup studied methods to identify the associated
momentum clumping using N-particle momentum correlations
\cite{Ran04a}.

We would like to study the granular source in configuration space.
Among the many different ways to study a quark-gluon plasma, intensity
interferometry (HBT interferometry) is the best experimental tool
to examine the space-time density distribution of the produced matter
\cite{Won94} -\cite{Bro04a}.  
It can be utilized to study the granular structure that occurs in a
first-order phase transition of the plasma.

It should be pointed out that without a careful study of the
phase-transition dynamics and post-transition evolution, it is not
known at present how much the granular density pattern of the phase
transition will remain and become detectable by HBT interferometry.
It also remains a subject of current research how early a
post-transition configuration the HBT interferometry really detects,
as it was shown by quantum treatments of the multiple scattering
process and the collective flow that HBT interferometry measures the
density distribution at a configuration earlier than the thermal
freeze-out configuration
\cite{Won03,Won04,Zha04a,Kap04}.  To continue our progress in the
search for the phase transition of a quark-gluon plasma, it is
reasonable to start with the working hypothesis that the density
fluctuations that occur during a first-order phase transition are so
large that some remnants of the granular droplet distribution remain
after the post-transition evolution, and these remnants form an initial
chaotic source of granular droplets. The assumed chaoticity of the
particle source then leads to the result that the correlation function
of two identical bosons in HBT interferometry contains information on
the space-time density of the source.  A granular density distribution
of the emitting chaotic source then distinguishes itself from other
density distributions and should reveal its characteristics in HBT
interferometry.

Suggestions to examine the granular structure in connection with the
phase transition of the quark-gluon plasma have been presented
previously \cite{Wit84}-\cite{Ran04a}.  Recent high-energy heavy-ion
measurements give a ratio of $R_{\rm out}/R_{\rm side} \sim 1$
\cite{PHE02,STA01}, which is contrary to most theoretical expectations
\cite{Pra03}-\cite{Mol04}.  A granular emitting source of droplets was
proposed to explain this HBT interferometry puzzle \cite{Zha04}.  The
suggestion was based on the observation that in the hydrodynamical
model \cite{Ris96}, the particle emission time scales with the radius
of the droplet.  Particles will be emitted earlier if the radius of
the droplet is smaller, as in a source of many droplets.  An
earlier emission time will lead to a smaller extracted HBT radius
$R_{\rm out}$.  On the other hand, $R_{\rm side}$ depends on the
distribution of droplet centers and is relatively independent of the
droplet size.  It increases with an increase in the width of the
droplet center distribution and the collective-expansion velocity of
the droplets.  As a result, the value of $R_{\rm out}$ can lie close
to $R_{\rm side}$ for a granular quark-gluon plasma source.  A direct
investigation on the granular density structure of the emitting source
is, however, needed to confirm the occurrence of a granular structure in
high-energy heavy-ion collisions.

Previously, Pratt $et~al.$ \cite{Pra92} studied the granular density
structure in HBT interferometry by considering pairs of identical
bosons in a distributed source in which the centers of the droplets
are distributed according to a spherical Gaussian distribution with a
standard deviation $R_0/\sqrt{2}$.  They calculated the correlation
function $C(|{\bf q}|)$ for a pair of identical bosons of relative
momentum ${\bf q}$ from such a distributed source with $R_0=4$ fm.
They found that there are differences in the magnitudes of the
correlation function $C(|{\bf q}|)$ at $|{\bf q}|$ greater than about
80 MeV/c in a granular structure.  They also found that the
correlation functions are relatively smooth functions of $|{\bf q}|$
even for small numbers of droplets (2 and 4).

The source considered by Pratt $et~al.$ in Ref.\ \cite{Pra92}
corresponds to a source in which the centers of the droplets have been
distributed first and identical bosons are then subsequently examined
from such a distributed source.  In actual dynamics with droplet
formation in a first-order phase transition \cite{Wit84}, one
envisages the formation of localized droplets in each single event
following the phase transition, and particles are emitted and evolve
from these localized granular droplets.  Although there can be a
distribution of the centers of the localized droplets over many
different heavy-ion collision events (of similar other global
characteristics), the centers of the droplets in each single collision
event can be localized.

Noting that the centers of the droplets in each single event can be
localized in space and time, we would like to consider a different way
to study the granular structure in HBT interferometry.  We propose
experiments to look at HBT interferometry for each event individually,
to study its associated density distribution.  There are many
advantages in the single-event HBT interferometry, if it can be carried
out with sufficient accuracy.  First, the granular structure is the
result of a large fluctuation of the density distribution in space and
time as it undergoes a first-order phase transition.  The large
fluctuation of the density in space and time is encoded into its
corresponding Fourier transform, whose absolute square gives the
correlation function of identical bosons.  By considering the
correlation function of all pairs of identical bosons from this single
event, one can obtain pertinent information on the density
distribution of the event.  One can even invert the correlation
function with an inverse Fourier transform to obtain an integral
equation for the source density distribution of the event.  Secondly,
as a result of the large fluctuation in the dynamics of a first-order
phase-transition, the granular structure of one event will be quite
different from the granular structure of another event, even though
many other global characteristics can be very similar.  By examining
the space-time structure of each event individually one can study the
large fluctuation of different events in the phase transition, which
is another characteristic of a first-order phase transition.  In other
words, there are both large fluctuations in a single event as well as
large fluctuations among different events in a first-order QCD phase
transition.  Finally, any averaging over a set of collision events to
obtain an average correlation function, as in the use of identical
boson pairs from different collision events in a `multi-event'
analysis or in considering bosons from a `distributed source', will
wash out the large fluctuations in the single-event correlation
function.  As a consequence, the large fluctuations of the correlation
function that are inherent in a first-order phase transition in single
events do not become prominent.  A single-event HBT analysis brings out
the prominent features of large fluctuations of the correlation
function.

At intermediate energies and the low-energy end of high-energy
heavy-ion collisions, it is difficult to use the single-event HBT
interferometry as the number of detected bosons is small in a single
event.  It becomes necessary to pick identical boson pairs from a
large number of events with similar global characteristics in a
`multi-event' analysis in order to provide sufficient statistics so as
to extract useful, but average, space-time characteristics concerning
this group of collisions. As a large number of events are included in
the sampling, the fluctuation of the correlation function that may be
present in the single-event HBT interferometry is suppressed by the
averaging procedure.

In a nearly head-on collision at very high energies, the number of
identical pions is of the order of a few thousand.  The number of
observed identical pions $n_\pi$ is a small fraction of this number.
For example, the number of identical pions detected in the STAR
Collaboration in the most central Au-Au collisions at RHIC is of the
order of a few hundred \cite{Ada04}.  Although the number of pairs of
identical pions in the event varies as $n_\pi(n_\pi-1)/2$, only a
small fraction of these pairs have relative momentum small enough to
be useful in a HBT analysis.  Clearly, whether or not a single-event
HBT measurement can be carried out remains to be tested.  If the
statistics of identical bosons turns out to be insufficient for such
an analysis using the present detectors in heavy-ion collisions at
RHIC, there remains the possibility of performing a single-event HBT
analysis at RHIC with detector upgrades or with heavy-ion collisions
at LHC.  It will also be of great interest to carry out a few-event HBT
analysis in future work both theoretically and experimentally to see
how the degrees of fluctuation changes as the number of events
increases.  The rate of the change of the fluctuations will provide
information on the underlying fluctuation in single-event and
event-to-event fluctuations in HBT interferometry.  The few-event
analysis will require a good theoretical understanding of the
single-event HBT interferometry.

We shall show that the correlation function of a granular density
distribution has large fluctuations, with maxima and minima at
locations which depend on the relative coordinates of the droplet
centers.  These local maxima and minima arise from the constructive
and destructive interference of identical bosons originating from two
different droplets.  Their interference pattern therefore depends on
the coordinates of the droplet centers. The occurrence of these maxima
and minima in single-event correlation functions is a good signature
for the granular structure and a first-order QCD phase transition.

It is desirable to take advantage of the relationship between the
density distribution and the correlation function to invert the
correlation function by Fourier transform to obtain an integral
equation for the density function.  Previously, using the Koonin-Pratt
formalism \cite{Kon78,Pra90}, Brown, Danielewicz, and their
collaborators studied methods to invert the correlation function by
expanding the two-particle source function in spherical harmonics
\cite{Bro97,Bro99,Bro04,Bro04a}.  For an irregular density
distribution as one encounters in granular droplets, an expansion in
terms of spherical harmonics is inadequate. We have developed a
general three-dimensional Fast Fourier Transform method to invert the
correlation function which will facilitate its application to HBT
interferometry, should accurate experimental single-event correlation
functions become available.

Using the property that the Fourier transform of the correlation
function is the folding of the density with itself, we shall prove
analytically that the Fourier transform of the correlation function of
a granular density distribution has maxima at spatial locations
governed by the spatial relative coordinates of the droplet centers.
The presence of this type of maxima in the Fourier transform of the
single-event correlation function provides another signature for the
granular structure of the density distribution.  We shall demonstrate
the presence of these maxima numerically by inverting the correlation
function with the three-dimensional Fast Fourier Transform (FFT)
method.  A comparison of these results with the exact analytical
solution shows the feasibility and the high degrees of accuracy of our
three-dimensional FFT method.

This paper is organized as follows.  In Section II, we briefly
summarize the basic relation between the density distribution and the
correlation function.  We give in Section III the correlation function
for a simple periodic structure and show that it possesses many
correlation function maxima.  In Section IV, we study the correlation
function of a general granular density distribution.  We find that the
correlation function exhibits large fluctuations.  We show that the
locations of the maxima and minima of the correlation function are
determined by the relative coordinates of the droplet centers.  In
Section V, we show that the Fourier transform of the correlation
function has maxima at locations which also depend on the relative
coordinates of the droplets centers.  The feasibility of the Fourier
transform method is demonstrated by numerical examples.  Section VI
concludes our discussions.  To facilitate the use of the Fourier
transform method using experimental single-event correlation
functions, (or functions that fit the experimental correlation
function), the method to invert a correlation function using the
3-dimensional Fourier transform is given in Appendix A. The method
to obtain the iterative solution of the density distribution from the
Fourier transform of the correlation function is given in Appendix B.

\section{Density Distribution and Correlation Function}

In a HBT interferometry measurement, one measures the correlation
function $P(k_1, k_2)$, the probability for the detection of one boson
of one 4-momentum $k_1=(k_{10},{\bf k}_1)$ in coincidence with the
detection of an identical boson of another 4-momentum $k_2$.  An
alternative representation of the correlation is given in terms of the
correlation function $C(k_1,k_2)$ defined as the ratio
$P(k_1,k_2)/P(k_1)P(k_2)$, where $P(k_i)$ is the probability of
detecting a boson of 4-momentum $k_i$.

Because of the symmetrization of the wave function of identical
particles, there is an interference of the two histories for two
identical bosons to propagate from two source points to two detection
points.  For a chaotic source, this interference leads to a relation
between the two-boson correlation function and the Fourier transform
of the source density $\rho(x;k_1,k_2)$ by \cite{Won94}-\cite{Bro04a}
\begin{eqnarray}
\label{ck1k2}
C(k_1,k_2)=1
+\bigl |\int dx e^{i(k_1-k_2)\cdot x} \rho(x;k_1,k_2) \bigr |^2,
\end{eqnarray}
where 
\begin{eqnarray}
\rho(x;k_1,k_2)=\frac{ \sqrt{f(k_1,x)f(k_2,x)} }
  {\sqrt{ \int dx_1 f(k_1,x_1) \int dx_2  f(k_1,x_1)}} ,
\end{eqnarray}
$f(k,x)$ is the Wigner function, and $x=({t,{\bf r}})$.  We shall
assume that the source is indeed chaotic so that Eq.\ (\ref{ck1k2})
holds.

The density function $\rho$ depends on the average momentum of the
pair of particles $P=(k_1+k_2)/2$ and the relative momentum
$q=k_1-k_2$.  We shall neglect final-state interactions and assume
that the density function $\rho$ is independent of $q$. Then the
average momentum $P$ becomes just a label for the correlation function
and the density function,
\begin{eqnarray}
C(q,P)=1
+\bigl |\int dx e^{iq\cdot x} \rho(x,P) \bigr |^2.
\end{eqnarray}
For brevity of notation, we shall leave the label $P$ implicit so that
the correlation function $C(q)$ and the density function $\rho(x)$ in
fact refers to those with a pair momentum $P$.  The label $P$ can be
re-introduced when it is needed.  In this implicit notation, the
correlation function $C(q)$ is related to the source density function
$\rho(x)$ by
\begin{eqnarray}
\label{cq}
C(q)=1 +\bigl |\int dx~ e^{iq\cdot x} \rho(x) \bigr |^2.
\end{eqnarray}  
We normalized the density function according to
\begin{eqnarray}
\label{nor}
\int ~dx~ \rho(x)= 1.
\end{eqnarray}

It is easy to show that the correlation function $C(q)$ is a real
function greater than unity and possesses an inversion symmetry,
\begin{eqnarray}
C(q)=C(-q).
\end{eqnarray}

Because of the simple relationship between $C(q)$ and $\rho(x)$ in
Eq.\ (\ref{cq}), there is a one-to-one mapping between the density
function $\rho(x)$ and the correlation function $C(q)$.  The
correlation function $C(q)$ will exhibit special properties associated
with the density distribution $\rho(x)$.  Conversely, information
concerning $\rho(x)$, including its fine structure, is encoded in
$C(q)$ so that direct information on $\rho(x)$ can, in principle, be
obtained by decoding $C(q)$ (see discussions in Section V).

In single-event HBT interferometry experiments, one measures
$P(k_1,k_2)$ in the same event and the correlation function
$C(k_1,k_2)$ is obtained by dividing the joint probability
$P(k_1,k_2)$ by $P(k_1)$ and $P(k_2)$.  Ideally, in the single event,
if the number of detected particles is large enough, the
single-particle probability distribution $P(k)$ can be well determined
using detected particles in the same event.  However, when one is
faced with low statistics in the total numbers of detected particles,
it may be necessary to obtain $P(k)$ from a larger set of events (with
similar global characteristics) to determine the correlation function
$C(k_1,k_2)$.

\section{An Example of a Regular Granular Source}

It is instructive to study many simple examples to find out the
characteristics of the correlation function of granular sources in HBT
interferometry.  We shall begin by considering a regular granular
structure.  While we do not expect the QCD phase transition to lead to
a completely regular granular structure, the remnants of the density
amplifications of the most rapidly amplified wavelengths that grow
predominantly during the first-order phase transition may remain, and
these amplified wavelengths may show up in the underlying gross
structure of an otherwise irregular granular pattern
\cite{Ran04}.  It is thus useful to study the correlation function of
a regular periodic granular structure.  The simplicity of the solution
also provides interesting insight into the nature of the correlation
function of granular density distributions.

Accordingly, we consider a spatially periodic granular source of the
form
\begin{eqnarray}
\label{per}
\rho(t,{\bf r})=f(t) \prod_{ j=0}^3 (\cos K_j x_j +1) \theta(R-r),
\end{eqnarray}
where the granular structure is characterized by an overall spatial
dimension $R$, and the wave numbers $K_j$ (or wavelengths
$d_j=2\pi/K_j$) with $j=1,2,3$.  The function $f(t)$ contains a
multiplicative normalization constant chosen to satisfy the
normalization condition (\ref{nor}).  The source is a periodic
granular distribution contained in a sphere of radius $R$.  The
Fourier transform of $\rho(x)$ is
\begin{eqnarray}
\tilde \rho (q_0,{\bf q}) 
&=& 
\tilde f (q_0) \int d{\bf r}~ e ^{-i {\bf q}\cdot {\bf r}}
 a({\bf r}) b({\bf r}),
\end{eqnarray}
where
\begin{eqnarray}
 \tilde f(q_0)= \int dt~ e ^{i q_0 t} f(t),
\end{eqnarray}
\begin{eqnarray}
 a({\bf r})= \prod_{j=1}^3 (\cos K_j x_j +1) ,
\end{eqnarray}
\begin{eqnarray}
 b({\bf r})
= \theta(R-r).
\end{eqnarray}
Using the folding theorem of Fourier transforms, we obtain
\begin{eqnarray}
 \int d{\bf r}~ e ^{-i {\bf q}\cdot {\bf r}}
 a({\bf r}) b({\bf r})
= \int \frac{d{\bf q}'}{(2\pi)^3} 
\tilde a({\bf q}') \tilde b({\bf q}-{\bf q}'),
\end{eqnarray}
where $\{\tilde a({\bf q}),\tilde b({\bf q}) \}$ are the Fourier
transforms of $\{a ({\bf r}),b ({\bf r}) \}$
respectively.  They are given explicitly by
\begin{eqnarray}
\tilde  a({\bf q})=(2\pi)^3 \prod_{j=1}^3
\left [ \delta (q_j ) + \delta(q_j-K_j)/2 + \delta(q_j+K_j)/2 \right ], 
\end{eqnarray}
\begin{eqnarray}
\tilde  b({\bf q})
=4 \pi R^3 \frac { j_1(qR)} {q R} .
\end{eqnarray}
From these results, we obtain
\begin{eqnarray}
\label{nor1}
\tilde \rho (q)= \tilde f(q_0)
\sum_{\lambda_1, \lambda_2, \lambda_3 = -1}^1  
c(\lambda_1) c(\lambda_2) c( \lambda_3) 
\frac{ j_1(|{\bf q} - {\bf Q}_{\lambda_1 \lambda_2 \lambda_3} |R)} 
     {|{\bf q} - {\bf Q}_{\lambda_1 \lambda_2 \lambda_3} |R},
\end{eqnarray}
where
\begin{eqnarray}
c(\lambda)= (1-|\lambda|/2), ~~~~~~~\lambda=-1,0,1,
\end{eqnarray}
\begin{eqnarray}
{\bf Q}_{\lambda_1 \lambda_2 \lambda_3}= 
\lambda_1 K_1 {\bf e}_1 + \lambda_2 K_2 {\bf e}_2 + \lambda_3 K_3 {\bf e}_3 ,
\end{eqnarray}
and $\{{\bf e}_1, {\bf e}_2, {\bf e}_3 \} $ are the unit vectors
along the $x$, $y$, and $z$ axes.  
The Fourier transform of the periodic density has maxima at
\begin{eqnarray}
\label{loc}
(q_1,q_1,q_3) = 
( \lambda_1 K_1 , \lambda_2 K_1, \lambda_3 K_3),
\end{eqnarray}
and each maximum can be labeled by the corresponding set of integers,
$\{ \lambda_1 \lambda_2 \lambda_3 \}$.  In units of $\tilde \rho(0)$,
the maximum of $\tilde \rho( {\bf q})$ at ${\bf q}=0$ is 1 (by
definition), and the maximum of $\tilde\rho({\bf q})$ at ${\bf
q}=(\lambda_1 K_1, \lambda_2 K_2, \lambda_3 K_3)$ is approximately
$c(\lambda_1) c(\lambda_2)c( \lambda_3)$.  The correlation function is
\begin{eqnarray}
C(q)=1+|\tilde f(q_0)|^2\left  [ 
\sum_{\lambda_1, \lambda_2, \lambda_3 = -1}^1  
c(\lambda_1) c(\lambda_2) c( \lambda_3) 
\frac{ j_1(|{\bf q} - {\bf Q}_{\lambda_1 \lambda_2 \lambda_3} |R)} 
     {|{\bf q} - {\bf Q}_{\lambda_1 \lambda_2 \lambda_3} |R}
\right ]^2.
\end{eqnarray}
In this case, $[C(q)-1]$ factorizes, and we can introduce a
three-dimensional correlation function $C({\bf q})$ defined by
\begin{eqnarray}
C(q)-1 = |\tilde f(q_0)|^2 [C({\bf q})-1].
\end{eqnarray}
For brevity of notation, we have used the same symbol $C$ for both the
three-dimensional and the four-dimensional correlation functions in the
above equation.  The ambiguity of the meaning of $C$ can be easily
resolved by context and by its argument.

The three-dimensional correlation function $C({\bf q})$ has maxima
values at the same locations as $\tilde \rho ({\bf q})$: at the origin
${\bf q}=0$ and at locations $q_j=\pm 2\pi|\lambda_j|/d_j$.  When $R
>> d$, the overlap of the individual distribution at various maxima is
small and we have approximately
\begin{eqnarray}
\label{ci}
C({\bf q}) \sim 1+
\sum_{\lambda_1, \lambda_2, \lambda_3 = -1}^1 
\left [  
c(\lambda_1) c(\lambda_2) c( \lambda_3) 
\frac{ j_1(|{\bf q} - {\bf Q}_{\lambda_1 \lambda_2 \lambda_3} |R)} 
     {|{\bf q} - {\bf Q}_{\lambda_1 \lambda_2 \lambda_3} |R} \right ] ^2.
\end{eqnarray}
The presence of this type of maxima in the three-dimensional
correlation function $C({\bf q})$ indicates that prominent
fluctuations are expected for a spatially periodic granular structure.

\section{General Granular Structure }

The occurrence of a first-order phase transition in the high
temperature environment of a high-energy heavy-ion collision will
generally lead to granular density distributions that are more general
than those discussed in the last section.  One can consider a general
granular density distribution of $N$ droplets of the type
\begin{eqnarray}
\rho (x) = A\sum_{j=1}^N  \rho_j(x-X_j),
\end{eqnarray}
where $\rho_j$ is the density distribution of the $j$-th droplet,
$X_j=(T_j,{\bf R}_j)$ is the localized space-time coordinate of the
center of the $j$-th droplet, and $A$ is a normalization constant such
that the total density $\rho(x)$ is normalized to unity as in Eq.\
(\ref{nor}).

We again assume that the source is chaotic in nature.  The two-boson
correlation function $C(q)$ is then related simply to the Fourier
transform of the source density.  The latter quantity can be easily
evaluated and found to be
\begin{eqnarray}
\tilde \rho (q)& =& \int dx ~ e^{i q \cdot x } \rho (x)\nonumber\\
&=& A \sum_{j=1}^N   e^{i q \cdot X_j} \tilde \rho_j (q),
\end{eqnarray}
where $\tilde \rho_j(q)$ is the Fourier transform of the density
distribution of the $j$-th droplet,
\begin{eqnarray}
\tilde \rho_j (q)& =& \int dx ~ e^{i q \cdot x } \rho_j (x).
\end{eqnarray}
The correlation function is then
\begin{eqnarray}
C(q)=1+\left |A \sum_{j=1}^N  e^{i q \cdot X_j} \tilde \rho_j (q)
\right |^2.
\end{eqnarray}
Thus, the correlation function of a granular source is related to the
absolute square of the coherent sum of the Fourier transforms of the
droplets modulated by the phase $e^{i q \cdot X_j}$, which depends on
the droplet center coordinate $X_j$.

We study the case where the space-time density distribution of a
droplet is given by a normalized Gaussian distribution with 
standard deviations $\sigma_j$ and $\tau_j$,
\begin{eqnarray}
\label{rhodis}
\rho_j(t,{\bf r}) = \frac{ e^{-{\bf r}^2/2\sigma_j^2}}
                       { (\sqrt{2\pi}\sigma_j )^3  }
                   \frac{ e^{-t^2/2\tau_j^2} }
                       { \sqrt{2\pi}\tau_j  }.
\end{eqnarray}
The Fourier transform of the density of a single droplet is 
\begin{eqnarray}
\tilde \rho_j (q)= \int dt\,  d {\bf r}\, 
e^{i q_0 t - i{\bf q} \cdot {\bf r} } \rho_j(t,{\bf r}) = 
e^{-\sigma_j^2{\bf q}^2/2}e^{-\tau_j^2q_0^2/2}. 
\end{eqnarray}
The total density of the granular source is the sum of the density of
the droplets,
\begin{eqnarray}
\label{csour}
\rho(x)=A \sum_{j=1}^N  
\frac{1}{(\sqrt{2\pi}\sigma_j)^3\sqrt{2\pi}\tau_j}
{\exp \left \{-\frac{({\bf r}-{\bf R}_j)^2}{2 \sigma_{j}^2}
              -\frac{(t-{T}_j)^2} {2 \tau_{j}^2}
\right \} }.
\end{eqnarray}
The correlation function for the Gaussian granular droplets is then
\begin{eqnarray}
C(q)=1+\left |A \sum_{j=1}^N  e^{i q \cdot X_j}  
 e^{-\sigma_j^2{\bf q}^2/2}
 e^{-\tau_j^2 q_0^2/2} 
\right |^2.
\end{eqnarray}
In general, the droplet lifetime $\tau_j$ may depend on the droplet
size parameter $\sigma_j$ (as in the hydrodynamical model
\cite{Ris96}), and $\sigma_j$ can be different for different droplets.

In order to get a clear insight into the most important features of
the correlation function of a granular structure, we consider the
simple case where the density distributions of all droplets are the
same so that $\sigma_j=\sigma_d$ and $\tau_j$=$\tau_d$ for all $j$.
In this simple case, $A=1/N$ and the correlation function can be
easily evaluated in terms of the positions of the droplet centers,
\begin{eqnarray}
\label{cqq}
C(q)=1+   \frac{ e^{-\sigma_d^2 {\bf q}^2-\tau_d^2 q_0^2} }{N^2}
\left | \sum_{j=1}^N  e^{iq_0T_j-i{\bf q }\cdot {\bf R}_j} 
\right |^2.
\end{eqnarray}
This leads to the simple result
\begin{eqnarray}
\label{eq31}
  C(q)=1+\frac {e^{-\sigma_d^2 {\bf q}^2-\tau_d^2 q_0^2} }
               {N^2}
\left [ N + 2 \sum_{j,k=1,j>k}^N \cos \{
   q\cdot (X_j - X_k)\}
 \right ].
\end{eqnarray}
Thus, the correlation function $C(q)$ has maxima at $q=0$ and $q_{\rm
max}\cdot{(X_j-X_k)}\sim 2n\pi$, with $n=1,2,3,...$.  It has minima at
$q_{\rm min}\cdot{(X_j-X_k)}\sim(2n-1)\pi$.

We shall consider further the simplified case in which the droplet
emission times $T_j$ are the same, then we have
\begin{eqnarray}
\label{eqT}
C(q)=1+   \frac{ e^{-\sigma_d^2 {\bf q}^2-\tau_d^2 q_0^2} }{N^2}
\left | \sum_{j=1}^N  e^{-i{\bf q }\cdot {\bf R}_j}
\right |^2.
\end{eqnarray}
The function $[C(q)-1]$ factorizes and we can introduce a
three-dimensional correlation function $C({\bf q})$ defined by
\begin{eqnarray}
\label{eq38}
C(q)-1 = e^{-\tau_d^2 q_0^2}[C({\bf q})-1].
\end{eqnarray}
We have again used the same symbol $C$ for both the three-dimensional
and the four-dimensional correlation function in the above equation.
From Eq.\ (\ref{cqq}), the three-dimensional correlation function
$C({\bf q})$ is given by
\begin{eqnarray}
\label{cqr}
C({\bf q})=1+\frac{e^{-\sigma_d^2 {\bf q}^2} } {N^2}
\left | \sum_{j=1}^N e^{-i{\bf q }\cdot {\bf R}_j}
\right |^2.
\end{eqnarray}
This leads to the simple result
\begin{eqnarray}
\label{cqrcos}
  C({\bf q})=1+\frac {e^{-\sigma_d^2 {\bf q}^2}}{N^2}
\left [ N + 2 \sum_{j,k=1,j>k}^N \cos \{{\bf q}\cdot ({\bf R}_j - {\bf R}_k)\}
 \right ].
\end{eqnarray}
The correlation function $C({\bf q})$ has a maximum at ${\bf q} =0$,
and $C({\bf q})|_{{\bf q}=0} =2$.  This maximum at ${\bf q}=0$ is a
common maximum as it occurs in all correlation functions of two
identical bosons.  From Eq.\ (\ref{cqrcos}) (and if correlation
function near the maxima and minima does not overlap), the correlation
function $C({\bf q})$ will have maxima approximately at
\begin{eqnarray}
\label{max}
{\bf q}_{\rm max}(jk,n)
\sim \left \{2n  \pi - \sin^{-1}
\left [\frac {(N+2)\sigma_d^2} {R_{jk}^2}\right ] \right \}
     \frac { {\bf R}_{jk} } {R_{jk}^2},
{\rm~~~~~for~~}j,k=1,...N, {\rm ~~and~~}j\ne k,
\end{eqnarray}
and $C({\bf q})$ will have minima
approximately at
\begin{eqnarray}
\label{min}
{\bf q}_{\rm min}(jk,n) 
\sim \left \{ (2n-1) \pi - \sin^{-1}
\left [\frac {(N-2)\sigma_d^2} {R_{jk}^2} \right ]  \right \}
     \frac { {\bf R}_{jk} } {R_{jk}^2},
{\rm~~for~~}j,k=1,...N, {\rm ~~and~~}j\ne k,
\end{eqnarray}
where $n=1,2,...$, ${\bf R}_{jk}={\bf R}_j-{\bf R}_k$, and
$R_{jk}=|{\bf R}_{jk}|$.  These maxima and minima correspond to
locations where the cosine function in Eq.\ (\ref{cqrcos}) has the
values close to +1 and $-1$ respectively.

It is easy to understand how these local maxima and minima of the
correlation function of a granular structure arise. The interference
of the histories of two bosons having a relative momentum ${\bf q}$
and originating from droplets $j$ and $k$ leads to a phase difference
of ${\bf q}\cdot ({\bf R}_j-{\bf R}_k)$. This phase interference gives
rise to the cosine function in Eq.\ (\ref{cqrcos}) and a local maximum
of the correlation function when the interference is constructive,
with a phase difference close to $2n\pi$.  It gives rise to a local
minimum when the interference is destructive, with a phase difference
close to $(2n-1)\pi$.

For each value of $n$, there are $N(N-1)$ maxima and $N(N-1)$ minima
of the correlation function $C({\bf q})$ of a granular structure, one
for each permutation of the pairs of droplets.  For each maximum
located at ${\bf q}_{\rm max}(jk,n)$, there is another maximum located
in the opposite direction at ${\bf q}_{\rm max}(kj,n)=-{\bf q}_{\rm
max}(jk,n).$ Similarly, ${\bf q}_{\rm min}(kj,n)=-{\bf q}_{\rm
min}(jk,n).$ The vectors ${\bf q}_{\rm max}(kj,n)$ and ${\bf q}_{\rm
min}(jk,n)$ are along the direction of the relative coordinate ${\bf
R}_j-{\bf R}_k$, and their magnitudes are inversely proportional to
the magnitudes of the relative coordinates.  The magnitude of the
maximum of the correlation function $[C({\bf q})-1]$, corresponding to
the pair of droplet centers $j$ and $k$, is approximately
$e^{-\sigma_d^2{\bf q}_{\rm max}^2(jk,n)}(N+2)/N^2$, and the magnitude
of the minimum is approximately $e^{-\sigma_d^2{\bf q}_{\rm
min}^2(jk,n)}(N-2)/N^2$.

The maxima and minima of the correlation function $C({\bf q})$ of a
granular structure, as given approximately by Eq.\ (\ref{max}) and
(\ref{min}), will maintain its distinct characteristics if they are
well separated.  However, these maxima and minima associated with
different pairs of droplets may be located in the vicinity of each
other.  They will merge to give rise to a more complicated pattern.
When the relative coordinates of many pairs of droplet centers are
approximately the same (as in the example of the periodic granular
structure discussed in Section III), the corresponding maximum of the
correlation function will be enhanced.  The shape of the correlation
function will be modified in the presence of overlapping maxima and
minima, to make the individual maximum and minimum less distinct.  It
is of interest to study some numerical examples to see whether these
interesting features of maxima and minima in the correlation function
manifest themselves.

If the coordinates of the droplet centers $\{ {\bf R}_j,~~ j=1,...N\}$
are known, then the correlation function $C({\bf q})$ can be evaluated
numerically.  We consider the center of a droplet ( $j$-th droplet,
say) in many different events to be distributed according to a
probability distribution $P({\bf R}_j)$,
\begin{eqnarray}
d P= P({\bf R}_j)\, d{\bf R}_j,
\end{eqnarray}
and assume $P({\bf R}_j)$ to be a normalized Gaussian distribution
with a standard deviation $\sigma_R$,
\begin{eqnarray}
\label{cdis}
d P = \frac{e^{-{\bf R}_j^2/2\sigma_R^2} } {(\sqrt{2\pi}\sigma_R )^3}
d{\bf R}_j.
\end{eqnarray}

In our numerical example, we randomly select the localized droplet
centers according to the distribution of Eq.\ (\ref{cdis}) with
$\sigma_R=4$ fm and take $\sigma_d=1.5$ fm for the standard deviation
of the droplet.  To avoid overlapping droplets, we require the
droplets be separated by a distance greater than the sum of their
root-mean-square radii, $2\sqrt{3}\sigma_d$.  After the positions of
the centers of the droplets are selected, we evaluate the correlation
function $C({\bf q})$ with Eq.\ (\ref{cqrcos}).  A sample result for
$N=4$ droplets is shown in Fig.\ 1, and another sample result for
$N=8$ droplets is shown in Fig. 2.  Figs.\ (1a) and (2a) show the
spatial configurations of the droplets; Figs.\ (1b) and (2b) give the
correlation functions $C({\bf q})$ as a function of $q_x$ and $q_y$
for $q_z=0$, Figs. (1c) and (2c) for $q_z=0.03$ GeV/c, and Figs. (1d)
and (2d) for $q_z=0.06$ GeV/c.  The momenta $q_x$ and $q_z$ in Figs.\
1 and 2 are in units of GeV/c.

One observes from Figs.\ 1 and 2 that there are prominent fluctuations
of the correlation function for a density distribution of localized
droplets.  The inversion symmetry $C(q_x,q_y,q_z)= C(-q_x,-q_y,-qz)$
is present for $q_z=0$ (Figs. (1b) and (2b)).  In addition to the
  maximum of $C(|{\bf q}|)$ at ${\bf q=0}$, there are maxima and
minima at various locations of ${\bf q}$.  The number of maxima for 8
droplets is greater than the number of maxima for 4 droplets.

The correlation functions for other configurations of 4 and 8
localized droplets exhibit similarly large fluctuations. On the
average, the smaller the number of droplets, the greater will be the
fluctuation.  The magnitude of the fluctuation decreases as the number
of droplets increases, as can be easily deduced from Eq.\
(\ref{cqrcos}).  The presence of this type of
maxima and minima of a single-event correlation function at many
relative momenta is a signature for a granular structure and a
first-order QCD phase transition.

Because of the large fluctuations in a first-order phase transition,
one expects that the locations of the centers of the droplets will be
quite different from event to event. The large differences in the
locations of the droplet centers in different events lead to large
differences in the locations of the maxima and minima and large
differences in the shapes of correlation functions (except for the
maximum at ${\bf q}$=0).

It is interesting to examine the correlation function when we average
over many events.  The correlation function $C({\bf q})$ depends on
the coordinates of the droplet centers, and the centers have a
distribution $P({\bf R}_j)$ in different events.  The average of the
correlation function over the different events, $\langle C({\bf
q})\rangle$, is defined as
\begin{eqnarray}
\langle C({\bf q}) \rangle =\int \prod_{j=1}^N d{\bf R}_j P({\bf R_j})
C({\bf q})
\Biggr / \int \prod_{j=1}^N  d{\bf R}_j P({\bf R_j}).
\end{eqnarray}
From Eq.\ (\ref{cqr}) and the Gaussian distribution of the droplet
centers (\ref{cdis}), we obtain
\begin{eqnarray}
\langle C({\bf q})\rangle=
1 + \frac{e^{-\sigma_d^2 {\bf q}^2}}{N}
+ \frac{e^{-\sigma_d^2 {\bf q}^2}}{N^2}
\sum_{j=1}^N \sum_{k=1,k\ne j}^N 
\int \frac{d {\bf R}_j d {\bf R}_k } {(\sqrt{2 \pi}\sigma_R)^6}
\exp \left \{-\frac{{\bf R}_j^2 +{\bf R}_k^2}{2\sigma_R^2} 
+ i {\bf q}\cdot( {\bf R}_j-{\bf R}_k) \right \},
\end{eqnarray}
and we get 
\begin{eqnarray}
\langle C({\bf q})\rangle=
1 + \frac{1}{N}e^{-\sigma_d^2 {\bf q}^2}
+ \frac{N-1}{N}e^{-(\sigma_d^2 + \sigma_R^2) {\bf q}^2},
\end{eqnarray}
which is identical to the results of Pratt $et~al.$ for a pair of
bosons from a distributed source \cite{Pra92}.  Thus, the results of
Pratt $et~al.$ \cite{Pra92} for a distributed source is the same as
the results of averaging the correlation function over many different
events.  The average correlation function is now a relatively smooth
function of $|{\bf q}|$, with only minor fluctuations even for only
two and four droplets.  The prominent fluctuations that are inherent
in single-event correlation functions involving the term $\cos {\bf
q}\cdot ({\bf R}_j-{\bf R}_k)$ in Eq.\ (\ref{cqrcos}) are not present.
The large fluctuations are now greatly suppressed by the averaging
procedure.

In order to bring out the salient features of the correlation
function, we have assumed in Eq.\ (\ref{eq31}) that the source
emission times $\{T_j\}$ for all droplets are the same.  This is a
reasonable assumption when the phase transition occurs over a short
duration of time.  The source function in Eq.\ (\ref{csour}) then
factorizes in spatial and temporal coordinates. The correlation
function $[C(q)-1]$ in Eq.\ (\ref{eq38}) also factorizes in $q_0$ and
${\bf q}$, with $C({\bf q})$ given by Eq.\ (\ref{cqrcos}).

On the other hand, if the phase transition occurs over a long
duration, then the emission times $\{ T_j \} $ can be different for
different droplets. The source function Eq.\ (\ref{csour}) cannot be
factorized as a product of spatial and temporal functions.
Consequently, the correlation function $[C(q)-1]$ cannot be factorized
and is given by the general result of Eq.\ (\ref{eq31}).  The
correlation function $C(q)$ has maxima at $q=0$ and $q_{\rm
max}\cdot{(X_j-X_k)}\sim 2n\pi$, with $n=1,2,3,...$.  It has minima at
$q_{\rm min}\cdot{(X_j-X_k)}\sim(2n-1)\pi$.  The maxima and minima of
the correlation function occur at relative momenta $q_0$ related to
the relative emission time coordinates, $T_j-T_k$.  The signature for
the granular structure remains distinct, if one can make accurate
measurements of the correlation function for different cuts in the
relative momenta $q_0$.

There are additional complications when one considers the internal
hydrodynamical motion and the collective motion of the droplets
relative to the center of mass.  The hydrodynamics of a QGP droplet
shows that the freeze-out radial coordinate in a QGP droplet is nearly
independent of time in a first-order phase transition, when the
initial energy density of the droplet is only slightly greater than
$\epsilon_c$, the QGP energy density at the critical temperature $T_c$
(see Fig. 4(b) of Ref. [32]).  As a droplet is presumably formed at
temperatures near the critical temperature with an energy densities
close to $\epsilon_c$, the assumption of the factorization of each
droplet source as a product of spatial and temporal functions is
reasonable for the case of a quark-gluon plasma with an initial
temperature slightly above $T_c$.

For the case of a quark-gluon plasma with a high initial temperature
much above $T_c$, the quark-gluon plasma will expand and cool.  It
will make a phase transition when it cools down to the critical
temperature with the subsequent formation of granular droplets, if the
phase transition is first-order in nature.  At the moment of phase
transition at $T_c$, the newly-formed droplets will acquire an
expansion velocity moving away from the center of mass of the system.
The factorization of the source function as a product of spatial and
temporal functions is not possible, and the magnitude of the
fluctuations of the correlation function may decrease.  We shall
investigate these effects on the correlation function in our future
work.

\section{ How to infer the density distribution from $C(q)$}

The correlation function $C(q)$ is related to the Fourier transform of
the density, $\tilde \rho(q)$.  Information on $\rho(x)$ is encoded in
$C(q)$.  If the correlation function $C(q)$ has been measured
experimentally as a function of its relative momentum coordinate $q$,
then a proper Fourier transform of the correlation function will
provide pertinent information on the density distribution $\rho(x)$.

We can obtain direct information on the density distribution $\rho(x)$
by decoding $C(q)$ in the following way.  From the correlation
function $C(q)$, one calculates $[C(q)-1]$, and one constructs the
Fourier transform of $[C(q)-1]$,
\begin{eqnarray}
\label{FT}
S(x)=\int \frac{dq}{(2\pi)^4} e^{-iq\cdot x}[C(q)-1].
\end{eqnarray}
In the discussion of the Fourier transform (or the inversion) of the
correlation function, the combination of the two terms in $[C(q)-1]$
always comes together.  For simplicity, we shall often use the term
`the Fourier transform (or the inversion) of the correlation function'
to mean `the Fourier transform (or the inversion) of $[C(q)-1]$'.

The function $S(x)$ is the two-particle `source function' in the
Koonin-Pratt formalism\cite{Kon78,Pra90} and the imaging method of
Brown, Danielewicz and their collaborators
\cite{Bro97,Bro99,Bro04,Bro04a}, for the special case for bosons
without final-state interactions. Much progress has been made in
obtaining a representation of this two-particle source function $S(x)$
in terms of spherical harmonics \cite{Bro97,Bro99,Bro04,Bro04a}.  For
an irregular density distribution and correlation function as one
encounters in granular droplets, an expansion of $S(x)$ in terms of
spherical harmonics will not be adequate. The irregularity of the
shape of the correlation function as shown in Figs.\ 1 and 2 calls for
a more general method.  The best method for inverting a general
correlation function without symmetry is to use Cartesian coordinates
in a three-dimensional Fourier transform.  We have developed
successfully a general three-dimensional Fast Fourier Transform method
to invert highly irregular three-dimensional correlation functions.
As described in details in Appendix A, our three-dimensional FFT
method consists of performing a sequence of one-dimensional cosine and
sine transforms in the three coordinate directions.  In each of the
cosine or sine transforms, we re-arrange the integral so that the
limits of the integration go from zero to infinity and the integrand
possesses the proper symmetric or antisymmetric reflectional symmetry,
for cosine or sine transform respectively.  We test our numerical
three-dimensional FFT method by applying it to invert a correlation
function of granular droplets for which results for the Fourier
inversion can be easily obtained analytically.  

In our analysis, we focus our attention on the source density function
$\rho(x)$ itself as it directly gives the space-time configuration of
the source.  From the relation between $C(q)$ and $\rho(x)$ as given
by Eq.\ (\ref{cq}), we get the integral equation for the source
density function $\rho(x)$,
\begin{eqnarray}
\label{rr}
{ S}(x)=\int   dx' ~\rho(x') ~\rho(x'+x).
\end{eqnarray}
We can prove that $S(x)$ possesses inversion symmetry
\begin{eqnarray}
{ S}(x)= S(-x).
\end{eqnarray}
The function ${ S}(x)$ is not the source density but is the folding of
the source density with itself.  To focus our attention on the source
density $\rho(x)$ and to emphasize the property of $S(x)$ as the
folding of $\rho(x)$ with $\rho(x)$, we can call the function $S(x)$
alternatively as `the folding function' of $\rho(x)$ in the discussion
of the source density $\rho(x)$, in addition to the name of `the
source function' in the Koonin-Pratt formalism \cite{Kon78,Pra90} and
imaging methods \cite{Bro97,Bro99,Bro04,Bro04a}.  The folding function
$S(x)$ is real and positive definite.  The same folding function
$S(x)$ is obtained whether one uses $e^{-iq\cdot x}$ or its complex
conjugate $e^{iq\cdot x}$ in the Fourier transform expression in Eq.\
(\ref{FT}).

We can illustrate the application of the folding function $S(x)$ with
the example of the chaotic source of $N$ Gaussian density droplets,
Eq.\ (\ref{csour}), studied in the last section.  For such a granular
source density $\rho(x)$, the folding function $S(x)$ can be obtained
analytically.  By carrying out the folding integration using Eq.\
(\ref{rr}), the folding function $S(x)$ can be easily found to be
\begin{eqnarray}
{ S}(x)=A^2
\sum_{j,\, k=1}^N 
 \frac{1}{(\sqrt{2\pi} \sigma_{jk})^3 \sqrt{2\pi} \tau_{jk} }
\exp \left \{
      -\frac{[{\bf r}-({\bf R}_j-{\bf R}_k)]^2}{2 \sigma_{jk}^2}
      -\frac{[t      -(T_j      -T_k      )]^2}{4 \tau_{jk}^2}
 \right  \}.
\end{eqnarray}
where $\sigma_{jk}^2=\sigma_j^2+\sigma_k^2$ and
$\tau_{jk}^2=\tau_j^2+\tau_k^2$.  The function $S(x)$ has maxima at
${\bf r}= ({\bf R}_j-{\bf R}_k)$ and $ t= (\tau_j-\tau_k)$, in addition
to the maxima at $x=0$.  For simplicity, we again assume that
$\sigma_j=\sigma_d$ and $\tau_j=\tau_d$ for all $j$.  The function
$S(x)$ for the granular droplets is simplified to be
\begin{eqnarray}
\label{sx}
{ S}(x)=\frac{1}{N^2(\sqrt{4\pi} \sigma_d)^3 \sqrt{4\pi} \tau_d }
\sum_{j,\, k=1}^N \exp \left \{
      -\frac{[{\bf r}-({\bf R}_j-{\bf R}_k)]^2}{4 \sigma_d^2}
      -\frac{[t      -(T_j      -T_k      )]^2}{4 \tau_d^2}
 \right  \}.
\end{eqnarray}
Thus, if the folding function near the maxima does not overlap, the
folding function $S(x)$ has maxima at locations governed by the
relative coordinates of the droplet centers.

We can consider the case when the droplets are all produced at the
same time, as for example when the droplets are produced at the moment
of phase transition.  Then the time $T_j$ can be taken to be the same
for all $j$. The four-dimensional folding function $S(x)$ factorizes
into a three-dimensional  part $S({\bf r})$ and a normalized Gaussian
distribution in time,
\begin{eqnarray}
S(x)=
S({\bf r}) \frac {e^{-t^2/4\tau_d^2}} {\sqrt{4\pi}\tau_d},
\end{eqnarray}
where we use the same symbol $S$ for the three- and four-dimensional
folding function.  The ambiguity of the meaning of $S$ can be easily
resolved by context and by its argument. The three-dimensional folding
function $S({\bf r})$ is
\begin{eqnarray}
\label{sx3d}
S({\bf r})
=\frac{1}{N^2(\sqrt{4\pi} \sigma_d)^3  }
\sum_{j,\, k=1}^N \exp \left \{
      -\frac{[{\bf r}-({\bf R}_j-{\bf R}_k)]^2}{4 \sigma_d^2}
 \right  \}.
\end{eqnarray}
If the folding function near the maxima does not overlap, the maxima of
the three-dimensional function $S({\bf r})$ are located at
\begin{eqnarray}
{\bf r}={\bf R}_j-{\bf R}_k, ~~~~~
~~~~~~~j,k=1,2,...N.
\end{eqnarray}
In Eq.\ (\ref{sx3d}) for $S({\bf r})$, there are $N$ terms with $j=k$,
and these terms contribute additively to the maxima at ${\bf r}=0$.
The height at the maxima at ${\bf r}=0$ is therefore $N$ times higher
than the maximum with $j \ne k$ located at the the relative
coordinates $ {\bf R}_j-{\bf R}_k$.  The occurrence of this type of
maxima in $S({\bf r})$, in addition to the maxima at ${\bf r}=0$,
provides another signature for a granular structure of the source and
a first-order phase transition of the quark-gluon plasma.

The density distribution we have considered, with both $\tau_j$ and
$T_j$ separately the same for all $j$, is a special case of those
density distributions whose spatial and time distributions can be
factorized: $\rho(x)=f(t) \rho({\bf r})$, where we use the same symbol
$\rho$ for the three-dimensional and the four-dimensional density
function.  For these factorizable density distributions,
\begin{eqnarray}
C(q)-1=|\tilde f(q_0)|^2[ C({\bf q})-1],
\end{eqnarray}
and
\begin{eqnarray}
 C({\bf q})-1=|\tilde \rho ({\bf q}) |^2,
\end{eqnarray}
where $\tilde f(q_0)$ and $\tilde \rho ({\bf q})$ are the Fourier
transforms of $f(t)$ and $\rho({\bf r})$, respectively.  Consequently,
the function $S(x)$ also factorizes and is given by
\begin{eqnarray}
S(x)= S({\bf r}) \int \frac{dq_0}{2\pi}e^{-iq_0 t} |\tilde f(q_0)|^2,
\end{eqnarray}
where the three-dimensional function $S({\bf r})$ is equal to
\begin{eqnarray}
\label{3d}
S({\bf r})=\int \frac{d {\bf q}}{(2\pi)^3}
e^{i{\bf q}\cdot {\bf r}} [C( {\bf q})-1].
\end{eqnarray}
It is of interest to demonstrate the feasibility and the accuracy of
the FFT method by using it to invert a correlation function and
comparing the inversion result with the analytical result.  We use the
numerical correlation function $C({\bf q})$ obtained in our previous
examples in Section IV (results as shown in Fig.\ 1 and 2) as input,
and carry out the three-dimensional FFT of the correlation function
$[C({\bf q})-1]$ to obtain $S({\bf r})$, as given in Eq.\
(\ref{3d}). The input correlation functions correspond to those of the
localized configurations of Figs.\ (1a) and (2a).

We show in Figs.\ 3 and 4 the results of the function $S({\bf r})$ at
$z=0, 1.94, 3.87,$ and 5.81 fm obtained by inverting the correlation
functions using the FFT method.  Fig.\ 3 gives $S({\bf r})$ for the
example of 4 droplets of Fig.\ 1.  Fig.\ 4 gives $S({\bf r})$ for the
example of 8 droplets of Fig.\ 2.  One observes that besides the
maxima at ${\bf r}=0$, the folding function $S({\bf r})$ has many
maxima at ${\bf r}={\bf R}_j-{\bf R}_k$ where $j,k=1,...N$ and $j\ne
k$. A granular structure shows up as having many maxima in the Fourier
transform of $[C({\bf q})-1]$, in addition to the maxima at ${\bf
r}=0$.  For the case with 4 droplets in Fig.\ 3, the maxima of $S({\bf
r})$ are quite distinctly exhibited.  For the case with 8 droplets,
the number of maxima increases and many maxima merge.  However, some
individual maxima remain distinctly separated as in Figs.\ (4b) and
(4c).

To study the shape of the function $S(x)$ in more detail, we consider
a cut at the plot of Fig.\ 3 at $x=0$, and plot $S(x=0,y,z)$ as a
function of $y$ for different values of $z$ in Fig. 5.  The results in
Fig.\ (5a) and (5b) have been obtained by using the Fast Fourier
Transform method for the example of 4 droplets of Fig.\ 1. Fig.\ (5a)
gives $S(0,y,z)$ in linear scale and (5b) in logarithmic scale.  One
sees clearly oscillations of the folding function $S(x)$ due to the
maxima at various ${\bf R}_j-{\bf R}_k$ locations.  A signature for
granular droplets is the presence of this type of maxima of the
Fourier transform of $[C({\bf q})-1]$ at various spatial locations.

We can assess the accuracy of inverting a numerical correlation
function with our Fast Fourier Transform method by comparing its
results with the exact analytical result as given by Eq.\
(\ref{sx3d}).  The exact analytical results of $S({\bf r})$ versus $y$
for $x=0$ and different $z$ values are shown in linear scale in Fig.\
(5c) and in logarithmic scale in Fig. (5d).  The results from the FFT
method match the exact analytical results with a very high degree of
accuracy, including the detailed shapes of the oscillations and the
values of $S({\bf r})$ down to the low density region of $y \sim \pm
30$ fm where $S({\bf r})$ is down by 6 orders of magnitude from its
maximum value at ${\bf r}=0$.  We have successfully developed an
accurate three-dimensional FFT method to invert a correlation function
$[C({\bf r})-1]$ to obtain its three-dimensional Fourier transform
$S({\bf r})$.

The folding function $S({\bf r})$ will be distorted as noises are
introduces into the correlation function.  The degree of distortion
will depend on the magnitude of the noise and it would be of interest
to see how well the folding function can be determined with noises
associated with experimental measurements.  The high degree of
accuracy in the FFT method makes it encouraging to apply it to
determine the folding function $S({\bf r})$ for the investigation of
the source density distribution $\rho({\bf r})$.

In order to facilitate the application of the three-dimensional
Fourier transform using the experimental single-event correlation
function $[C({\bf q})-1]$ (or perhaps functions that fit the
experimental correlation function), we give the detailed steps of how
the three-dimensional Fourier transform can be evaluated in Appendix
A.  The computer program to carry out the three-dimensional Fast
Fourier Transform to obtain $S({\bf r})$ from $[C({\bf q})-1]$ can
also be obtained from the authors upon request.  Brown $et~al.$ have
pointed out that in practical applications, when the experimental
errors are large, it is important to include the error uncertainties
into the equation for the inversion of the correlation function
\cite{Bro97,Bro99,Bro04,Bro04a}. Brown \cite{Bro99} also pointed out 
that when one applies the FFT transform to experimental correlation
functions, one should take care to treat the experimental error in the
measurement by filtering out the noise, and the best method is one in
which the errors of the measurement are included in the inversion
method.

It is worth emphasizing that Eq.\ (\ref{rr}), which connects the
Fourier transform of the correlation function $[C(q)-1]$ to the
density function $\rho(x)$, is a general result.  It can be used to
obtain other density distributions, in addition to the granular
density distribution discussed here. Thus, if the correlation function
$C(q)$ is experimentally determined, one can first evaluate the
Fourier transform of $[C(q)-1]$, which gives the function $S(x)$.  The
integral equation Eq.\ (\ref{rr}) can then be used to determine the
density distribution $\rho(x)$ by algebraic methods. In the
three-dimensional case, one can discretize the integral equation Eq.\
(\ref{rr}) as
\begin{eqnarray}
\label{iter}
{ S}(i,j,k)= \Delta x \Delta y \Delta z 
\sum_{i',j',k'=1}^N \rho (i+i', j+j', k+k') \rho(i',j',k'),
\end{eqnarray}
where $x=i\Delta x$, $y=j \Delta y$, and $z=k \Delta z$.  We consider
only the region of $S(x)$ and $\rho(x)$ inside the box of $i,j,k =1$
to $N$ and assume that they are zero outside the box.

With the determination of $S({\bf r})$ from a given experimental
correlation function by the FFT method, one can in principle solved the
above equation to obtain the source density distribution $\rho({\bf
r})$.  One can for example solve the above equation by iteration.  In
the first iterative step, one uses a initial guessed density
distribution $\rho^{(0)}({\bf r})$ as one of the two density functions in
Eq.\ (\ref{iter}). The equation is then a linear equation of $\rho({\bf r})$
and can be easily solved.  The steps to obtain the solution is
described in Appendix B.  The solution $\rho({\bf r})$ can then be
substituted into Eq.\ (\ref{iter}) to replace one of the two $\rho$
functions and to continue the iteration.  Clearly, the iterative
solution $\rho({\bf r})$ will be the solution of Eq.\ (\ref{rr}) or
(\ref{iter}) if the iteration converges.  The successes of the
iterative solution will probably depend on a good initial guessed
solution.  It will be of great interest to test how this iterative
procedure may be used to find a density distribution $\rho({\bf r})$ for a
given experimental correlation function $C({\bf q})$.  It is necessary to
investigate how one can guarantee positive-definite solutions of
$\rho({\bf r})$ in such procedure.  Future development to search for methods
to solve the integral equation Eq.\ (\ref{rr}) [or (\ref{iter})] for
$\rho(x)$ from a given $S(x)$ will be of great interest.

\section{ Conclusions and Discussions}

Recent experiments at RHIC provide ample evidence for a dense matter
produced in high-energy heavy ion collisions.  Is the produced dense
matter the quark-gluon plasma?  An unambiguous identification of the
produced matter as a quark-gluon plasma requires the observation of
the phase transition from the new form of matter to known hadronic
matter.  

The signature for a phase transition depends on the order of the phase
transition.  Witten and many workers noted previously that a granular
structure of droplets occurs in a first-order QCD phase transition,
and the observation of the granular structure can be used as a
signature for a first-order QCD phase transition
\cite{Wit84}-\cite{Ran04a}.  HBT interferometry is the best
experimental tool to examine the space-time density distribution of
the produced matter. It can therefore be utilized to study the
granular structure that occurs in a first-order phase transition of
the plasma.

In the dynamics following a first-order QCD phase transition, the
evolved matter will react chemically and thermally.  It is not known
how much the granular density pattern of the phase transition will
remain to make it detectable by HBT interferometry.  It has been argued
in conventional theory that HBT interferometry measures the density
distribution of the hadron matter at thermal freeze-out, as the
re-scattering of bosons is assumed to lead to a chaotic configuration.
This traditional assumption is subject to question as it was, however,
pointed out recently that the propagation of bosons in the
re-scattering process should be investigated in a quantum description
\cite{Won03,Won04,Zha04a,Kap04}. Upon using the Glauber theory to describe 
the scattering process, it was found that HBT interferometry measures
the initial chaotic density distribution modified by absorption and
collective flow \cite{Won03,Won04}.  The HBT interferometry may be
sensitive to the density distribution that occurs earlier than the 
thermal freeze-out configuration.  If the initial density fluctuation
is large, a substantial density fluctuation of the granular pattern may
remain to make it detectable by HBT interferometry.

Whatever the theoretical predictions on the possible evolution of the
produced matter may be, it is ultimately an experimental question
whether a granular density distribution that occurs at the moment of
the phase transition may subsequently render itself detectable by the
available experimental tool of HBT interferometry.

We propose new ways to detect a granular density structure using
the single-event HBT interferometry.  If it can be carried out with
sufficient accuracy, the single-event HBT interferometry can reveal the
density distribution in each single event.  It can also reveal large
fluctuations in the density distribution from event to event, as is
expected in a first-order phase transition.

We carry out our analysis with many examples of granular sources.  We
found that a granular structure is characterized by large fluctuations
of the single-event correlation function. A single-event correlation
function has maxima and minima at relative momenta that depend on the
relative coordinates of the droplet centers.  The presence of this
type of maxima and minima of the single-event correlation function can
be used as the signature for a granular structure and the first-order
QCD phase transition of the quark-gluon plasma.

If an experimental single-event correlation function is complete and
accurate enough, another very simple method to search for a granular
structure is to take the Fourier transform of the correlation
function.  The Fourier transform of the correlation function leads to
the folding of the source density with itself.  This Fourier transform
possesses maxima at spatial coordinates governed by the relative
coordinates of the droplet centers.  The occurrence of this type of
maxima in the Fourier transform, in addition to the maxima at ${\bf
r}=0$, is another signature for granular droplets and a first-order
quark-gluon plasma phase transition.

In the present analysis, we have focused our attention on the
single-event HBT interferometry to emphasize the maximum fluctuations
in the correlation function of a granular structure.  It will be of
great interest to study a few-event HBT analysis both theoretically
and experimentally. The few-event HBT analysis will be necessary for
practical reasons, if the number of boson pairs in a single event is
not large enough to provide sufficient statistics.  The few-event HBT
analysis will also be needed to understand the degrees of fluctuation
from event to event.  One wishes to find out whether the fluctuations
in few-event HBT correlation measurements contain a single-event
component that is beyond statistical fluctuations.  The rate of the
change of the degrees of fluctuation as the number of events increases
will provide information on the underlying fluctuations in the
single-event and event-to-event HBT interferometry.  The successful
development of the single-event or few-event HBT interferometry in
high-energy heavy-ion collisions will open up a vast vista for future
exploration.

In the present investigation, we have considered idealized situations
in order to bring out the most important features of the signature for
a granular structure.  It will be of great interest to examine in
future work how the signature discussed here may be affected when some
of our simplifying assumptions are modified.  The determining factor
for the occurrence of maxima and minima in the correlation function
$C({\bf q})$ is the interference of two histories for two bosons to
propagate from two source points in different droplets to the
detecting points.  This interference involves a phase difference which
depends on the radius vector joining the two droplets.  If this
underlying factor of interference leading to large fluctuations of the
correlation function remains important even after modifying our
simplifying assumptions, then many of the gross features obtained here
will not be greatly modified.  

The fluctuations arising from a granular structure described in the
present idealized theoretical investigation will be blurred by
experimental statistical fluctuations due to the limited number of
experimental counts.  Whether or not the relevant signal can be
recovered in the presence of experimental statistical fluctuations
remains to be tested.  Clearly, the smaller the number of droplets,
the greater is the signal and the greater will be the probability of
its observation in the presence of statistical fluctuations.  It will
be of great interest to carry out a theoretical simulation to see what
are the minimum droplet size and the largest droplet number a given
experimental arrangement may be able to detect.

Much work remains to be done both experimentally and theoretically to
investigate this interesting topic on the signature of the phase
transition of the quark-gluon plasma.  It will be of interest to study
theoretically effects of the collective expansion of the sources and
the droplets, effects of fluctuations of the size of the droplets,
effects of absorption of the bosons on its way to the detector,
effects of the momentum dependences of the density distributions, and
other interesting questions in connection with the signature of the
granular structure.

\begin{acknowledgments}

The authors would like to thank T. Awes, V. Cianciolo, T. D. Lee, and
J. Randrup for helpful and stimulating discussions.  This research was
supported by the National Natural Science Foundation of China under
Contract No.10275015 and by the Division of Nuclear Physics, US DOE,
under Contract No. DE-AC05-00OR22725 managed by UT-Battelle, LLC.
\end{acknowledgments}

\appendix
\section{Evaluation of the three-dimensional Fourier transform of the 
correlation function $[C({\bf q})-1)]$}

From Eq.\ (\ref{3d}), we have
\begin{eqnarray}
S({\bf r})&=&\int \frac{d{\bf q}}{(2\pi)^3} 
e^{-i{\bf q}\cdot {\bf r}}[C({\bf q})-1]
\nonumber\\
&=& \int \frac{d{\bf q}}{(2\pi)^3} \{ \cos({\bf q}\cdot{\bf r})
+ i \sin({\bf q}\cdot{\bf r}) \}
 [C({\bf q})-1].
\end{eqnarray}
The imaginary part in the above integration vanishes as $C(-{\bf
q})=C({\bf q})$.  It is only necessary to evaluate the real part of
the Fourier transform.  The folding function $S({\bf r})$ is
therefore
\begin{eqnarray}
S({\bf r})=\int \frac{d{\bf q}}
{(2\pi)^3} & &\cos(q_x x + q_y y + q_z z)~R({\bf q}),
\end{eqnarray}
where $R({\bf q})=C({\bf q})-1$.  Expanding the cosine function, we obtain
\begin{eqnarray}
\label{exp}
S({\bf r}) =\int \frac{d{\bf q}}{(2\pi)^3}& &[\cos q_x x \cos q_y y
\cos q_z z -\cos q_x x \sin q_y y \sin q_z z\nonumber\\ & & ~~ -\sin
q_x x \cos q_y y \sin q_z z -\sin q_x x \sin q_y y \cos q_z z ]~ R({\bf
q}).
\end{eqnarray}
In the standard numerical subroutines such as those in the Fast
Fourier Transform package of DFFTPACK \cite{FFT}, the cosine transform
of an even function of $G(q)$ is usually approximated by
\begin{eqnarray}
\label{cos}
\tilde G(x)=\int_0^\infty dq ~\cos(qr) G(q)
=\Delta q \sum_{j=1}^N \cos\{ (j-1)(k-1)\pi/N \} G(q),
\end{eqnarray}
where $x=(k-1)\Delta x$, $q=(j-1)\Delta q$, and $(\Delta x)(\Delta
q)=\pi/N$ .  The right-hand side quantities are then calculated by the
one-dimensional cosine Fast Fourier Transform subroutine of the
package.  Similarly, the sine transform of an odd function is usually
approximated by
\begin{eqnarray}
\label{sin}
\tilde G(x)=\int_0^\infty dq ~\sin(qr) G(q)
=\Delta q \sum_{j=1}^N \sin \{ jk\pi/(N+1) \} G(q),
\end{eqnarray}
with $x=k \Delta x$ and $q=j\Delta q$ and $\Delta x \Delta q
=\pi/(N+1)$.

The Fourier integral of the sine and cosine function in Eq.\
(\ref{exp}) can be cast into the standard form of Eqs. (\ref{cos}) and
(\ref{sin}) in the FFT package of DFFTPACK by noting that
\begin{eqnarray}
\int_{-\infty}^{\infty} dq_x~ \cos q_x x ~g(q_x,q_y,q_z)
=\int_0^{\infty}dq_x~\cos q_x x\, [ g(q_x,q_y,q_z) + g(-q_x,q_y,q_z) ],
\end{eqnarray}
and similarly
\begin{eqnarray}
\int_{-\infty}^{\infty} dq_x~ \sin q_x x ~ g(q_x,q_y,q_z)
=\int_0^{\infty} dq_x~ \sin q_x x \,[ g(q_x,q_y,q_z) - g(-q_x,q_y,q_z) ].
\end{eqnarray}
Using the above results, the four terms inside the square bracket in
Eq.\ (\ref{exp}) lead to four contributions to $S({\bf r})$,
\begin{eqnarray}
\label{beg}
S({\bf r}) =[A_1({\bf r})-A_2({\bf r}) -A_3({\bf r}) -A_4({\bf r})
]/(2\pi)^3,
\end{eqnarray}
where
\begin{eqnarray}
\label{a9}
A_1(x,y,z)=\int_0^\infty  dq_z~ \cos{q_z z}[F_1(x,y,q_z)+F_1(x,y,-q_z)],
\end{eqnarray}
\begin{eqnarray}
F_1(x,y,q_z)=\int_0^\infty  dq_y~ \cos{q_y y}[E_1(x,q_y,q_z)+E_1(x,-q_y,q_z)],
\end{eqnarray}
and
\begin{eqnarray}
E_1(x,q_y,q_z)=\int_0^\infty  dq_x~ \cos{q_x x}
\left  [ R(q_x,q_y,q_z)+ R(-q_x,q_y,q_z) \right ].
\end{eqnarray}
The term $A_2$ that contributes to $S(x)$ can be obtained similarly as
\begin{eqnarray}
A_2(x,y,z)=\int_0^\infty  dq_z~ \sin{q_z z}[F_2(x,y,q_z)-F_2(x,y,-q_z)],
\end{eqnarray}
\begin{eqnarray}
F_2(x,y,q_z)=\int_0^\infty  dq_y~\sin{q_y y}[E_2(x,q_y,q_z)-E_2(x,-q_y,q_z)],
\end{eqnarray}
and
\begin{eqnarray}
E_2(x,q_y,q_z)=\int_0^\infty dq_x~ \cos{q_x x}
\left [ R(q_x,q_y,q_z)+ R(-q_x,q_y,q_z) \right ].
\end{eqnarray}
In a similar way, the term $A_3(x,y,z)$ is given by 
\begin{eqnarray}
A_3(x,y,z)=\int_0^\infty dq_z~\sin{q_z z}[F_3(x,y,q_z)-F_3(x,y,-q_z)],
\end{eqnarray}
\begin{eqnarray}
F_3(x,y,q_z)=\int_0^\infty dq_y~ \cos{q_y y}[E_3(x,q_y,q_z)+E_3(x,-q_y,q_z)],
\end{eqnarray}
and
\begin{eqnarray}
E_3(x,q_y,q_z)=\int_0^\infty dq_x~\sin({q_x x})
\left [R(q_x,q_y,q_z)- R(-q_x,q_y,q_z) \right ].
\end{eqnarray}
Finally,  $A_4(x,y,z)$ is given by
\begin{eqnarray}
A_4(x,y,z)=\int_0^\infty dq_z~\cos{q_z z}[F_4(x,y,q_z)+F_4(x,y,-q_z)],
\end{eqnarray}
\begin{eqnarray}
F_4(x,y,q_z)=\int_0^\infty dq_y~\sin{q_y y}[E_4(x,q_y,q_z)-E_4(x,-q_y,q_z)],
\end{eqnarray}
and
\begin{eqnarray}
\label{end}
E_4(x,q_y,q_z)=\int_0^\infty dq_x~\sin {q_x x}
\left [R(q_x,q_y,q_z)- R(-q_x,q_y,q_z) \right ].
\end{eqnarray}
The right-hand sides of the above equations (\ref{a9})-(\ref{end})
are now in the form of the sine and cosine integrals of Eqs.\
(\ref{cos}) and (\ref{sin}), for which standard FFT subroutines can be
applied.  With the above relations, the folding function $S({\bf r})$
can be easily evaluated using subroutines in standard Fast Fourier
Transform packages.

Incidentally, we have shown how we can obtain the 3-dimensional
Fourier transform for a function that is symmetric with respect to the
inversion of its coordinates, for which the imaginary part of the
Fourier transform vanishes.  For other general functions without such
a symmetry, the imaginary Fourier component involving $\sin ({\bf
q}\cdot {\bf r})$ does not vanish.  We can expand the function $\sin (
q_x x +q_y y + q_z z )$ as in Eq.\ (\ref{exp}) and use techniques
similar to those in Eqs.\ (\ref{beg})-(\ref{end}) to get the imaginary
part of the Fourier transform.

\section{Solution of the discretized integral equation}

We wish to obtain an iterative solution of the discretized integral
equation (\ref{iter}) for $\rho({\bf r})$, with a given $S({\bf r})$.
As the density function $\rho({\bf r})$ is zero outside the box of
$\{i,j,k =1$ to $N\}$.  the summation in Eq.\ (\ref{iter}) can be
limited to density functions inside the box.  Consequently, Eq.\
(\ref{iter}) contains fewer and fewer numbers of unknown variables of
$\rho$ as the indices $i$, $j$, or $k$ of $S(i,j,k)$ increases.  We
can choose our starting point to be the linear equation containing
only one variable.  The equation can be easily solved.  The variables in
the subsequent set of linear equation can be solved in sequence.
Similar procedures can be carried out in two and three dimensions.  We
shall give the detail procedures below for the one-dimensional case to
indicate how the iterative solution can be obtained.

We seek an iterative solution
of $\rho(i)$ satisfying 
\begin{eqnarray}
\label{B1}
S(i)/\Delta x =
\sum_{i'=1}^N \rho (i+i') \rho^{(0)}(i'),
\end{eqnarray}
where $\rho^{(0)}(i')$ is either a guessed solution or the
solution of the previous iteration.

Because the density function is zero outside the region of $i<1$ and $i>N$,
the set of equations of (\ref{B1})  are
\begin{eqnarray}
S(N)=0,
\end{eqnarray}
\begin{eqnarray}
\label{1st}
S(N-1)/\Delta x=\rho(N) \rho^{(0)}(1), 
\end{eqnarray}
\begin{eqnarray}
\label{2nd}
S(N-2)/\Delta x=\rho(N-1) \rho^{(0)}(1) + \rho(N) \rho^{(0)}(2),
\end{eqnarray}
\begin{eqnarray}
S(N-3)/\Delta x=\rho(N-2) \rho^{(0)}(1) 
+ \rho(N-1) \rho^{(0)}(2) + \rho(N) \rho^{(0)}(3), ....
\end{eqnarray}
\begin{eqnarray}
S(1)/\Delta x=\rho(1) \rho^{(0)}(1) + \rho(2) \rho^{(0)}(2)+... + \rho(N) \rho^{(0)}(N).
\end{eqnarray}
Eq.\ (\ref{1st}) contains only a single unknown, $\rho(N)$, and can be
solved in terms of the other known quantities.  Knowing the value of
$\rho(N)$, Eq.\ (\ref{2nd}) can be solved for $\rho(N-1)$, and so on.
In this way, the whole array of $\rho(i)$ can be determined.  The
above method can be easily generalized to calculate the iterative
solution of the 3-dimensional density function $\rho({\bf r})$ from
$S({\bf r})$.  It will be necessary to normalize the density solution
$\rho({\bf r})$ after each iterative step to ensure that the final
solution has the proper normalization.

\section{Figures}

\vspace*{1.0cm}
\begin{figure}
\includegraphics[scale=0.60]{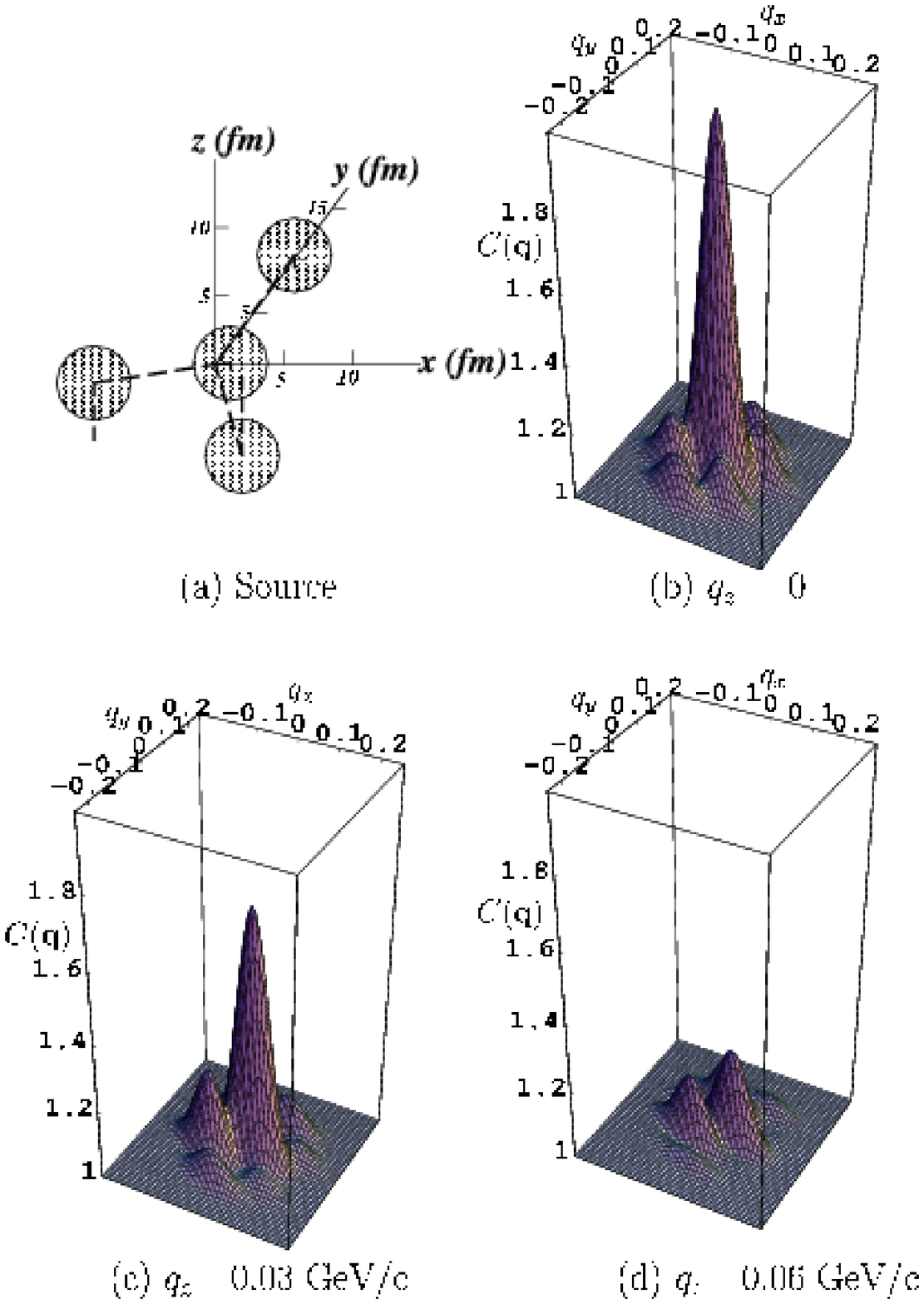}
\vspace*{-2.0cm}
\caption{ (a) A sample spatial configuration of four droplets.
We use dashed lines to join the droplet centers and the origin to
show the locations of the droplets.  We use vertical dashed lines to
indicate the projection of the droplet centers onto the $x-y$
plane. (b) Two-boson correlation function $C(q_x,q_y,q_z)$ for
$q_z=0,$ (c) for $q_z=0.030$ GeV/c, and (d) $q_z=0.06$ GeV/c.  The
quantities $q_x$ and $q_y$ are in units of GeV/c. }
\end{figure}

\newpage
\begin{figure}
\includegraphics[scale=0.60]{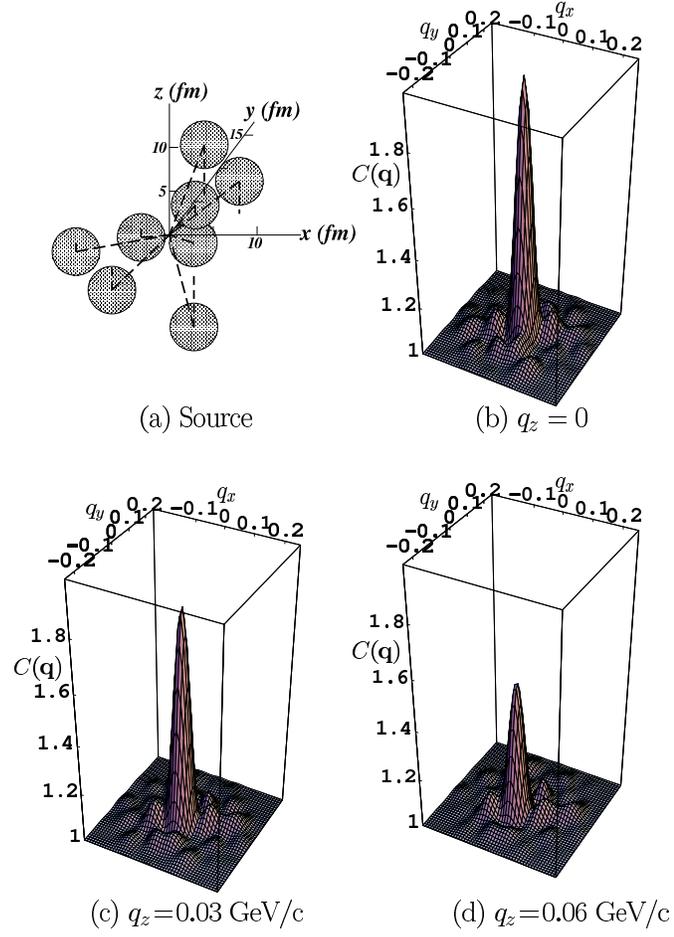}
\vspace*{-1cm}
\caption{\label{fig:f2} (a) A sample configuration of eight droplets.
We use dashed lines to join the droplet centers and the origin to
show the locations of the droplets.  We use vertical dashed lines to
indicate the projection of the droplet centers onto the $x-y$ plane.
(b) Two-boson correlation function $C(q_x,q_y,q_z)$ as a function of
$q_x$ and $q_y$ at $q_z=0,$ (c) $q_z=0.03$ GeV/c, and (d) $q_z=0.06$
GeV/c.  The quantities $q_x$ and $q_y$ are in units of GeV/c.}
\end{figure}

\newpage
\begin{figure}
\includegraphics[scale=0.60]{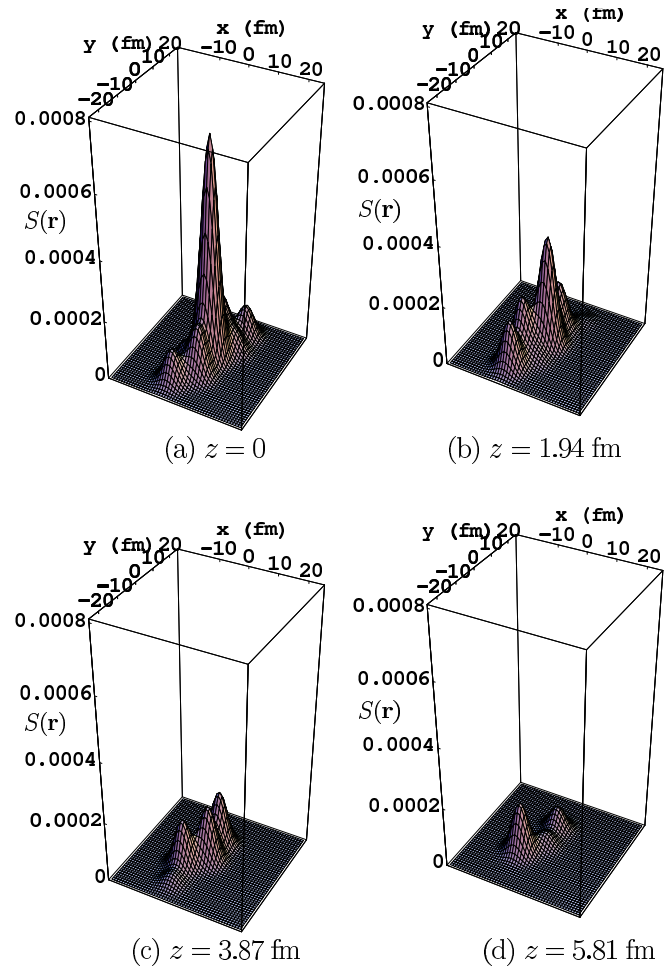}
\vspace*{-3cm}
\caption{\label{fig:f3} 
(a) The folding function $S(x,y,z)$ for 4 droplets in units of
fm$^{-3}$ for the droplet configuration of Fig. 1.  Fig.\ (3a) is for
$S({\bf r})$ at $z=0$, (3b) for $S({\bf r})$ at $z=1.94$ fm, (3c) for
$S({\bf r})$ at $z=3.87$ fm, and (3d) for $S({\bf r})$ at $z=5.81$ fm.
}
\end{figure}

\newpage

\begin{figure}
\includegraphics[scale=0.60]{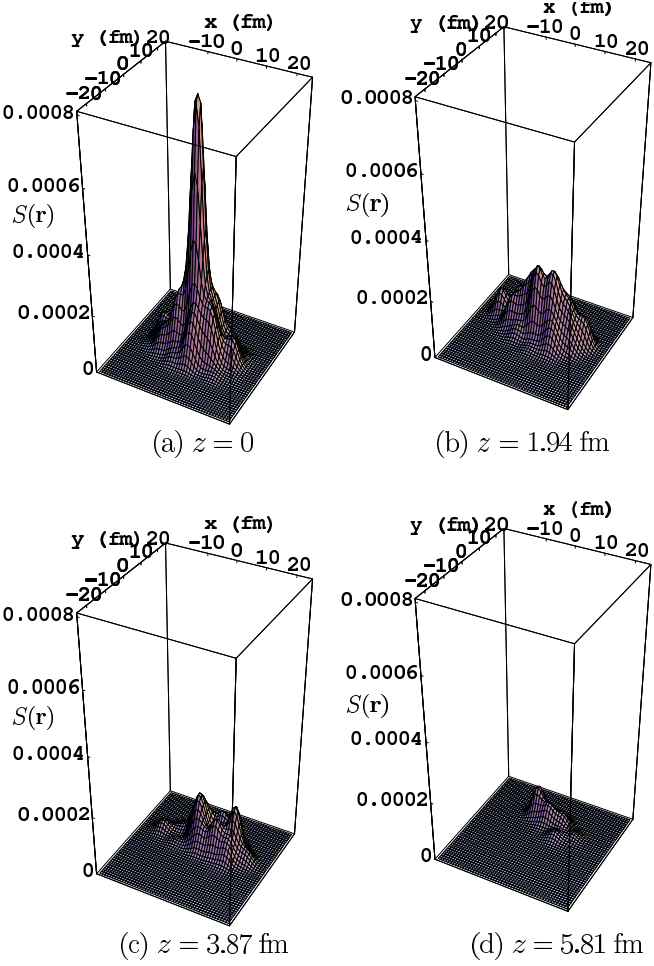}
\vspace*{-3cm}
\caption{ (a) The function $S(x,y,z)$ 
for 8 droplets in units of fm$^{-3}$ for the droplet distributions of
Fig. 2.  Fig.\ (4a) is for $S({\bf r})$ at $z=0$, (4b) for $S({\bf
r})$ at $z=1.94$ fm, (4c) for $S({\bf r})$ at $z=3.87$ fm, and (4d)
for $S({\bf r})$ at $z=5.81$ fm.  }
\end{figure}

\newpage

\begin{figure}
\includegraphics[scale=0.60]{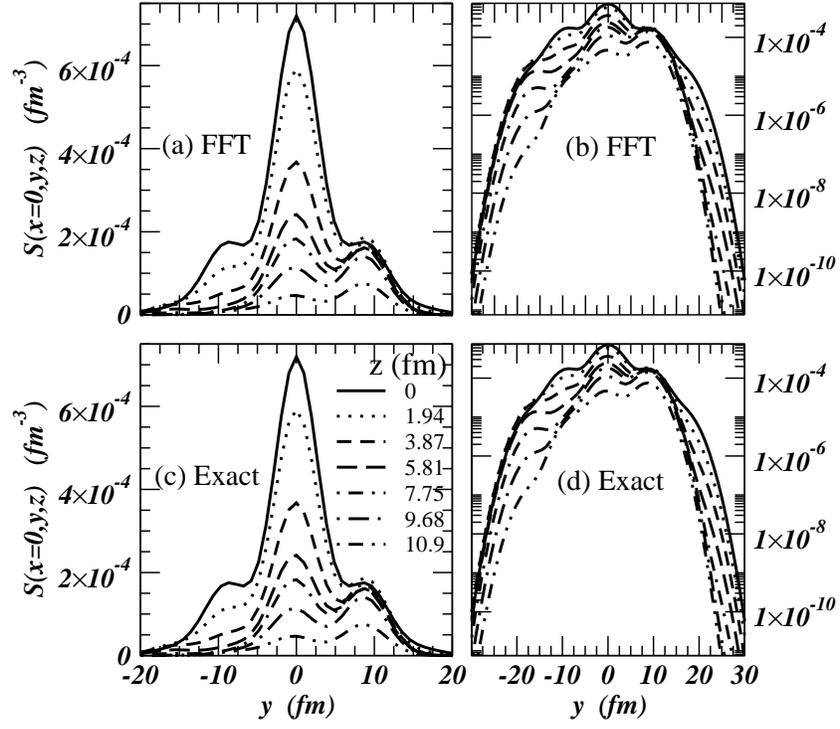}
\vspace*{-0.0cm}
\caption
{ The function $S(x=0,y,z)$ versus y for various values of $z$ for the
4 droplets examples of Figs.\ 1 and 3. The results from the FFT Method
are given in linear scale in Fig.\ (5a) and in logarithmic scale in
Fig.\ (5b).  The results from the exact analytical solution of Eq.\
(\ref{sx3d}) are given in linear scale in Fig.\ (5c) and in logarithmic
scale in Fig.\ (5d).}
\end{figure}

\end{document}